\documentclass[aps,prl,twocolumn,superscriptaddress,showpacs,floatfix,nofootinbib]{revtex4-1}
\pdfoutput=1

\usepackage{graphicx}
\usepackage{amssymb}   
\usepackage{amsmath}
\usepackage{amsopn}
\usepackage{tabularx}
\usepackage{pdftexcmds}
\usepackage[T1]{fontenc}

\begin{document}


\title{Measurement of Reactor Antineutrino Oscillation Amplitude and Frequency at RENO}

%

\affiliation{Institute for Universe and Elementary Particles, Chonnam National University, Gwangju 61186, Korea          }
\affiliation{Institute for High Energy Physics, Dongshin University, Naju 58245, Korea                     }
\affiliation{GIST College, Gwangju Institute of Science and Technology, Gwangju 61005, Korea         }
\affiliation{Institute for Basic Science, Daejeon 34047, Korea     }
\affiliation{Department of Physics, KAIST, Daejeon 34141, Korea          }
\affiliation{Department of Physics, Kyungpook National University, Daegu 41566, Korea          }
\affiliation{Department of Physics and Astronomy, Seoul National University, Seoul 08826, Korea }
\affiliation{Department of Fire Safety, Seoyeong University, Gwangju 61268, Korea              }
\affiliation{Department of Physics, Sungkyunkwan University, Suwon 16419, Korea                }

\author{G. Bak}
\affiliation{Institute for Universe and Elementary Particles, Chonnam National University, Gwangju 61186, Korea          }
\author{J. H. Choi}
\affiliation{Institute for High Energy Physics, Dongshin University, Naju 58245, Korea                     }
\author{H. I. Jang}
\affiliation{Department of Fire Safety, Seoyeong University, Gwangju 61268, Korea              }
\author{J. S. Jang}
\affiliation{GIST College, Gwangju Institute of Science and Technology, Gwangju 61005, Korea         }
\author{S. H. Jeon}
\affiliation{Department of Physics, Sungkyunkwan University, Suwon 16419, Korea                }
\author{K. K. Joo}
\affiliation{Institute for Universe and Elementary Particles, Chonnam National University, Gwangju 61186, Korea          }
\author{K. Ju}
\affiliation{Department of Physics, KAIST, Daejeon 34141, Korea                  }
\author{D. E. Jung}
\affiliation{Department of Physics, Sungkyunkwan University, Suwon 16419, Korea                }
\author{J. G. Kim}
\affiliation{Department of Physics, Sungkyunkwan University, Suwon 16419, Korea                }
\author{J. H. Kim}
\affiliation{Department of Physics, Sungkyunkwan University, Suwon 16419, Korea                }
\author{J. Y. Kim}
\affiliation{Institute for Universe and Elementary Particles, Chonnam National University, Gwangju 61186, Korea          }
\author{S. B. Kim}
\affiliation{Department of Physics and Astronomy, Seoul National University, Seoul 08826, Korea }
\author{S. Y. Kim}
\affiliation{Department of Physics and Astronomy, Seoul National University, Seoul 08826, Korea }
\author{W. Kim}
\affiliation{Department of Physics, Kyungpook National University, Daegu 41566, Korea          }
\author{E. Kwon}
\affiliation{Department of Physics and Astronomy, Seoul National University, Seoul 08826, Korea }
\author{D. H. Lee}
\affiliation{Department of Physics and Astronomy, Seoul National University, Seoul 08826, Korea }
\author{H. G. Lee}
\affiliation{Department of Physics and Astronomy, Seoul National University, Seoul 08826, Korea }
\author{Y. C. Lee}
\affiliation{Department of Physics and Astronomy, Seoul National University, Seoul 08826, Korea }
\author{I. T. Lim}
\affiliation{Institute for Universe and Elementary Particles, Chonnam National University, Gwangju 61186, Korea          }
\author{D. H. Moon}
\affiliation{Institute for Universe and Elementary Particles, Chonnam National University, Gwangju 61186, Korea          }
\author{M. Y. Pac}
\affiliation{Institute for High Energy Physics, Dongshin University, Naju 58245, Korea                     }
\author{Y. S. Park}
\affiliation{Institute for Universe and Elementary Particles, Chonnam National University, Gwangju 61186, Korea          }
\author{C. Rott}
\affiliation{Department of Physics, Sungkyunkwan University, Suwon 16419, Korea                }
\author{H. Seo}
\affiliation{Department of Physics and Astronomy, Seoul National University, Seoul 08826, Korea }
\author{J. W. Seo}
\affiliation{Department of Physics, Sungkyunkwan University, Suwon 16419, Korea                }
\author{S. H. Seo}
\affiliation{Department of Physics and Astronomy, Seoul National University, Seoul 08826, Korea }
\author{C. D. Shin}
\affiliation{Institute for Universe and Elementary Particles, Chonnam National University, Gwangju 61186, Korea          }
\author{J. Y. Yang}
\affiliation{Department of Physics and Astronomy, Seoul National University, Seoul 08826, Korea }
\author{J. Yoo}
\affiliation{Institute for Basic Science, Daejeon 34047, Korea     }
\affiliation{Department of Physics, KAIST, Daejeon 34141, Korea                  }
\author{I. Yu}
\affiliation{Department of Physics, Sungkyunkwan University, Suwon 16419, Korea                }

\collaboration{The RENO Collaboration}

%
%

\begin{abstract}

The RENO experiment reports more precisely measured values of $\theta_{13}$ and $|\Delta m_{ee}^2|$ using $\sim$2\,200 live days of data. The amplitude and frequency of reactor electron antineutrino ($\overline{\nu}_e$) oscillation are measured by comparing the prompt signal spectra obtained from two identical near and far detectors. In the period between August 2011 and February 2018, the far (near) detector observed 103\,212 (850\,666) $\overline{\nu}_e$ candidate events with a background fraction of 4.8\% (2.0\%). A clear energy and baseline dependent disappearance of reactor $\overline{\nu}_e$ is observed in the deficit of the measured number of $\overline{\nu}_e$. Based on the measured far-to-near ratio of prompt spectra, we obtain $\sin^2 2 \theta_{13} = 0.0896 \pm 0.0048({\rm stat}) \pm 0.0047({\rm syst})$ and $|\Delta m_{ee}^2| =[2.68 \pm 0.12({\rm stat}) \pm 0.07({\rm syst})]\times 10^{-3}$~eV$^2$.
\end{abstract}


\maketitle

%

  The smallest neutrino mixing angle $\theta_{13}$  is firmly measured by the reactor $\overline{\nu}_e$ disappearance \cite{RENO, DB, DC}. It establishes a complete picture of neutrino oscillations among three flavors.
Due to a rather large $\theta_{13}$ value, a next round of neutrino experiments~\cite{future-exp} are under preparation or consideration for determining the CP violating phase in the leptonic sector and the neutrino mass ordering.
A more precise measurement of $\theta_{13}$ by a reactor experiment will greatly improve the CP phase determination. 
Reactor experiments with a baseline distance of $\sim$1 km can also determine an effective squared mass difference $\Delta m_{ee}^2 \equiv \cos^2 \theta_{12}\Delta m_{31}^2 + \sin^2 \theta_{12} \Delta m_{32}^2$~\cite{Parke} using the oscillation frequency in the $\overline{\nu}_e$ survival probability $P$ ~\cite{Petcov}. The probability is given by
\begin{eqnarray}
P &  \approx 1  &  - \sin^2 2 \theta_{13} \sin^2 \Delta_{ee}   \nonumber   \\
   &                   &  - \cos^4 \theta_{13}\sin^2 2\theta_{12} \sin^2 \Delta_{21}
,~\label{eqn:Papprx}
\end{eqnarray}
where $\Delta_{ij} \equiv 1.267 \Delta m_{ij}^2 L/E$, $E$ is the $\overline{\nu}_e$ energy in MeV, and $L$ is the distance  between the reactor and detector in meters.

The first measurement of $|\Delta m_{ee}^2|$ by RENO was reported based on the rate, spectral and baseline information of reactor $\overline{\nu}_e$ disappearance using $\sim$500 live days of data \cite{RENO-spect1, RENO-spect2}. In this Letter, we present more precisely measured values of $\theta_{13}$ and $|\Delta m_{ee}^2|$ using $\sim$2\,200 live days of data.
The systematic uncertainty in the measurement is reduced due to better understanding of backgrounds and increased data size.

The RENO experiment has been in data-taking since August, 2011.
Identical near and far $\overline{\nu}_e$ detectors are placed 294 and 1383~m, respectively, from the center of six reactor cores of the Hanbit Nuclear Power Plant so that a ratio measurement may cancel out possible correlated systematic uncertainties between them. The plant consists of six pressurized water reactors, each with maximum thermal output of 2.8~GW$_{\rm th}$, that are situated in a linear array spanning 1.3 km with equal spacings. The reactor flux-weighted baseline is 410.6~m for the near detector and 1445.7~m for the far detector. 

A reactor $\overline{\nu}_e$ is detected through the inverse beta decay (IBD) interaction, $\overline{\nu}_e + p \rightarrow e^+  + n$, in hydrocarbon liquid scintillator (LS) with 0.1\% gadolinium (Gd). A prompt signal from the positron annihilation releases energy of 1.02 MeV as two $\gamma$-rays in addition to the positron kinetic energy. The neutron after thermalization is captured by Gd with a mean delayed time of $\sim$26 $\mu$s and produces several $\gamma$-rays with the total energy of $\sim$8 MeV. 
The RENO LS is made of linear alkylbenzene with fluors. A Gd-carboxylate complex was developed for the best Gd loading efficiency into LS and its long term stability \cite{RENO-GdLS}.  

Each RENO detector consists of a main inner detector (ID) and an outer veto detector (OD). The ID is contained in a cylindrical stainless steel vessel that houses two nested cylindrical acrylic vessels \cite{RENO-acrylic}. The innermost acrylic vessel holds 16.5 tons of Gd-doped LS as a neutrino target, and is surrounded by a $\gamma$-catcher region with a 60 cm thick layer of undoped LS inside an outer acrylic vessel. Outside the $\gamma$-catcher is a 70~cm thick buffer region filled with mineral oil. Light signals emitted from particles are detected by 354 low background 10 in. photomultiplier tubes (PMTs) \cite{RENO-PMT} that are mounted on the inner wall of the stainless steel container. The 1.5~m thick OD region is filled with highly purified water, and equipped with 67 10 in. PMTs mounted on the wall of the concrete OD vessel.
More detailed description of the RENO detectors can be found in Refs. \cite{RENO-spect2, RENO-proposal}.

An event energy is given by the total charge ($Q_{{\rm tot}}$) in photoelectrons (p.e.) that is collected by the PMTs within $-$100 to $+$50~ns and corrected for gain and charge collection variations using the neutron capture peak energies. 
The absolute energy of a prompt event ($E_{\rm p}$) is determined by the corrected $Q_{{\rm tot}}$ using a charge-to-energy conversion function obtained from various source calibration samples and neutron capture samples.
Detailed discussion on the energy calibration can be found in Refs. \cite{RENO-spect1, RENO-spect2}.

The observed $Q_{{\rm tot}}$ is reduced by $\sim$15\% of the initial operation value due to decrease of LS attenuation length, and by $\sim$10\% due to unplugged flashing PMTs. The decreased attenuation length is caused by loose air-tightening around the detector chimney region and most likely introducing oxygen and moisture into the LS.  
The attenuation length remains unchanged after careful air shielding with nitrogen gas. 	
A nonuniform charge response in the detector volume is developed by the decreased attenuation length. A spatial correction using the delayed energy peak is applied to recover a unifrom charge response.

In this measurement we use 2\,193.04 (1\,807.88) live day data in the far (near) detector, taken in the period between August 2011 and February 2018. The near data sample in the period of January to December 2013 is not used because of detection inefficiency caused by an electrical noise coming from an uninterruptible power supply. A small amount of $^{252}$Cf was accidentally introduced into both detectors during detector calibration in October 2012. Most of multiple neutron events coming from $^{252}$Cf contamination are eliminated by multiplicity requirements.

IBD candidate events are obtained by selection criteria including a time coincidence requirement of 2 to 100 $\mu$s between a promptlike event and a delayedlike event of neutron capture by Gd.
A detailed description of the selection criteria is given in Refs. \cite{RENO-spect1, RENO-spect2}. 
Some of them are modified to remove more backgrounds and reduce their uncertainties as follows.
Firstly, the timing veto and muon visible energy ($E_{\mu}$) criteria are optimized for additional reduction of cosmogenic backgrounds, mainly coming from $\beta$-n emitters from cosmic-muon induced $^9$Li/$^8$He isotopes.
Events associated with the muon are rejected if they are within a 1\,000 ms (800 ms, 500 ms, 100 ms)  window following a cosmic muon of $E_{\mu} > 1.5$ GeV (1.3$-$1.5~GeV, 1.1$-$1.3~GeV, 0.85$-$1.1~GeV) for the far detector, or within a 800 ms (300 ms, 200 ms, 50 ms)  window following a cosmic muon of $E_{\mu} > 1.6$ GeV (1.4$-$1.6~GeV, 1.3$-$1.4~GeV, 1.1$-$1.3~GeV) for the near detector.
The improved muon-veto requirement reduces the remaining $^9$Li/$^8$He background rate by 36.5\% (38.9\%) in the far (near) detector with an additional signal loss of 7.2\% (4.6\%).
Secondly, a tighter spatial correlation requirement of $\Delta R < 2.0$~m is imposed for additional reduction of accidental backgrounds where $\Delta R$ is the distance between the prompt and delayedlike events.
The tighter spatial requirement reduces the remaining accidental background rate by 53.0\% (63.1\%) in the far (near) detector. 
Thirdly, stringent multiplicity requirements are applied to remove more $^{252}$Cf contamination background events in the far detector where the contamination is higher than the near detector.   
IBD candidates are rejected (i) if there is another subsequent IBD pair within the 1~s interval, (ii) if any ID triggers other than those associated with a delayed event occur within 800~$\mu$s from its prompt event, or (iii) if they are accompanied by a prompt event of  $E_{\rm p} > 3$~MeV within a 30~s window and a distance of 50~cm.
After applying the requirements, 99.9\% of the $^{252}$Cf contamination background events are eliminated. The remaining $^{252}$Cf contamination background rates are estimated to be 0.43$\pm$0.04 (0.08$\pm$0.02) per day in the far (near) detector.

The muon and multiplicity timing veto requirements are applied differently to the near and far detectors. 
The IBD signal loss due to the tighter requirements is 31.252$\pm$0.045\% (39.671$\pm$0.005\%) for the far (near) detector.
The background rate is reduced to 70.9\% (54.3\%) of the previously measured value \cite{RENO-spect2, RENO-proposal} for the far (near) detector. The background uncertainty is reduced from 7.3\% (4.7\%) to 4.3\% (3.1\%) for the far (near) detector.

Applying the selection criteria yields 103\,212 (850\,666) IBD candidates with $1.2 <E_{\rm p} <8.0$~MeV in the far (near) detector. 
In the final data samples, the remaining backgrounds are either uncorrelated or correlated IBD candidates between the prompt and delayedlike events. An accidental background comes from random association  of prompt and delayedlike events. Correlated backgrounds are fast neutrons from outside of the ID, $\beta$-$n$ emitters from cosmic-muon induced $^9$Li/$^8$He isotopes, and $^{252}$Cf contamination. 
The remaining background rates and spectral shapes are obtained from control data samples \cite{RENO-spect1, RENO-spect2}.
The total background rates are estimated to be $2.24\pm0.10$ and  $9.53\pm0.28$ events per day for far and near detectors, respectively. The total background fraction is $4.76\pm0.20$\% in the far detector, and $2.03\pm0.06$\% in the near detector. 
The observed IBD and background rates are summarized in Table \ref{tab:Event_rate}. 

\begin{table}[hbt]
 \caption{Measured IBD and estimated background rates with  $1.2 <E_p <8.0$~MeV, given in per day.}
 \begin{center}
 \begin{tabular*}{0.48\textwidth}{@{\extracolsep{\fill}} l r r }
 \hline \hline
     Detector              & Near       &  Far      \\
  \hline
IBD rate                     &   $470.53\pm0.51$ &  $47.06\pm0.15$ \\
After background subtraction   &   $461.00\pm0.58$  &   $44.82\pm0.18$ \\
Total background rate  &   $ 9.53\pm0.28$ &  $2.24\pm0.10$ \\
Live time (days)    &    1807.88        &   2193.04    \\
    \hline
  Accidental rate    &    $2.54\pm0.03$   &  $0.46\pm0.01$  \\
  $^9$Li/$^8$He rate &    $5.10\pm0.27$   &  $0.98\pm0.08$   \\
  Fast neutron rate  &    $1.81\pm0.02$   &  $0.37\pm0.01$   \\
  $^{252}$Cf contamination rate &   $0.08\pm0.02$   &  $0.43\pm0.04$   \\ 
  \hline \hline
  \end{tabular*}
 \end{center}

 \label{tab:Event_rate}

 \end{table}  

The prompt energy difference between the near and far detectors contributes to the uncorrelated systematic uncertainties associated with a relative measurement of spectra at two detectors and is estimated by comparing the energy spectra of various $\gamma$-ray sources using the charge-to-energy conversion functions. The uncorrelated energy scale difference is found to be less than 0.15\% from all the calibration data.

The average detection efficiency of the near and far detectors is 76.47$\pm$0.16\% with an uncorrelated systematic uncertainty of 0.13\%. Main contributions to the uncorrelated uncertainty come from different efficiencies between the two detectors associated with Gd-capture fraction and delayed energy requirement. The uncorrelated systematic uncertainty on the Gd capture fraction is estimated as 0.1\% due to the difference of Gd concentration between two detectors. The uncertainty on the delayed energy requirement is estimated as 0.05\% from the delayed energy uncertainty of 0.15\%. A fractional error of the detection efficiency is 0.21\% to be used as the uncertainty of the far-to-near detection efficiency ratio. A detailed description of the detection efficiency can be found in Ref. \cite{RENO-spect2}.

The expected rates and spectra of reactor $\overline{\nu}_e$ are calculated for the duration of physics data-taking by taking into account the varying thermal powers, fission fractions of four fuel isotopes, energy release per fission, fission spectra, and IBD cross sections \cite{Vogel, Feilitzsch, Hahn, Declais, Mueller, Huber, Kopeikin}. 
The total uncorrelated systematic uncertainty of reactor flux is estimated as 0.9\%. The total correlated uncertainty of reactor flux is 2.0\% and is cancelled out in the far-to-near ratio measurement.

We observe a clear deficit of the measured IBD rate in the far detector with respect to the expected one, indicating the reactor $\overline{\nu}_e$ disappearance. 
Using the deficit information only, a rate-only analysis obtains $\sin^2 2 \theta_{13}$ = 0.0874 $\pm$ 0.0050(stat) $\pm$ 0.0054(syst), where the world average value of $|\Delta m_{ee}^2|= (2.56 \pm0.05) \times 10^{-3}$~eV$^2$ is used \cite{PDG}.
The total systematic error of $\sin^2 2 \theta_{13}$ is reduced from 0.0068 to 0.0054, mostly due to the decreased background uncertainty, relative to the previous measurement~\cite{RENO-spect1, RENO-spect2}, while the statistical error is significantly reduced from 0.0091 to 0.0050.

\begin{figure}[hbt]
\begin{center}
\includegraphics[width=0.47\textwidth]{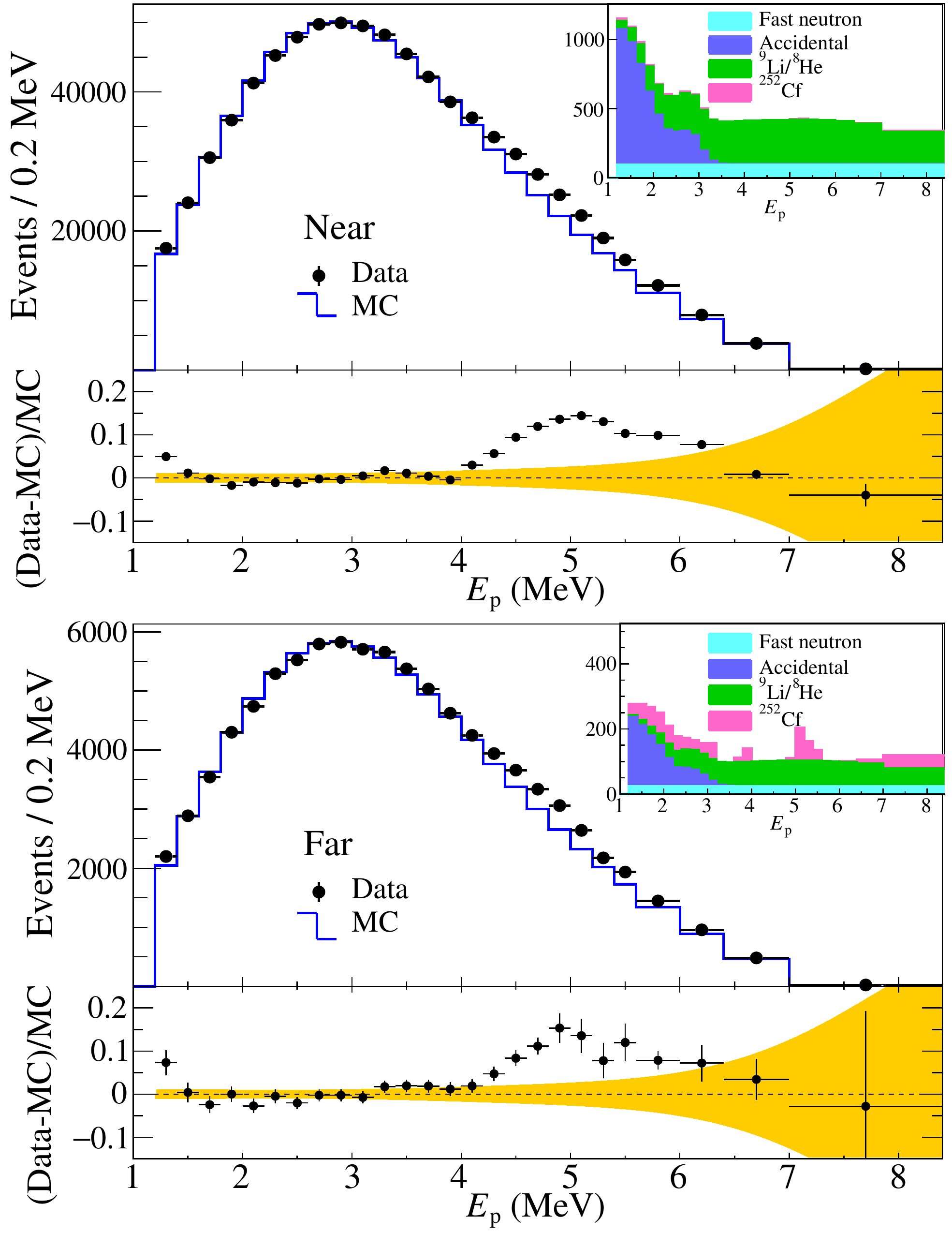}
\caption{Spectral shape comparison of observed and expected IBD prompt events in the near and far detectors. 
The observed spectra are obtained from subtracting the remaining background spectra as shown in the insets.
The expected distributions are obtained from the best-fit oscillation results that are applied to the no-oscillation MC spectra. The deviation from the expectation near 5 MeV is larger than the uncertainty of an expected spectrum (shaded band) from the reactor antineutrino model\cite{Mueller, Huber}.}
\label{fig:IBD-shape-comp}
\end{center}
\end{figure}

Figure \ref{fig:IBD-shape-comp} shows a shape comparison between the observed IBD prompt spectrum after background subtraction and the prediction from a reactor $\overline{\nu}_e$ model \cite{Mueller, Huber} and the best-fit oscillation results. 
The fractional difference between data and prediction is also shown in the lower panel.
A clear discrepancy between the observed and MC predicted spectral shapes is found in the region of 5~MeV in both detectors. For the spectral shape comparison, the MC predicted spectrum is normalized to the observed one in the region excluding $3.6 < E_{\rm p} < 6.6$~MeV. This observation suggests needs for reevaluation and modification of the current reactor $\overline{\nu}_e$ model~\cite{Mueller, Huber}. 

We observe a clear energy dependent disappearance of reactor $\overline{\nu}_e$ in the far detector.
Even with the unexpected structure around 5 MeV, the oscillation amplitude and frequency can be determined from a fit to the measured far-to-near ratio of IBD prompt spectra because of its cancellation in the ratio measurement. The relative measurement using identical near and far detectors makes the method insensitive to the correlated uncertainties of expected reactor $\overline{\nu}_e$ flux and spectrum as well as detection efficiency.
For determination of $|\Delta m_{ee}^2|$ and $\theta_{13}$ simultaneously, a $\chi^2$ with pull parameter terms of systematic uncertainties is constructed using the spectral ratio measurement and is minimized by varying the oscillation parameters and pull parameters as described in Refs.~\cite{RENO-spect1, RENO-spect2}.

The systematic uncertainty sources are embedded by pull parameters with associated systematic uncertainties. 
The pull parameters allow variations from the expected far-to-near ratio of IBD events within their corresponding systematic uncertainties.
The uncorrelated reactor-flux uncertainty is 0.9\%, the uncorrelated detection ratio uncertainty is 0.21\%, the uncorrelated energy scale uncertainty is 0.15\%, and the background uncertainty is 5.61\% and 3.26\% for far and near detectors, respectively.

The best-fit values obtained from the rate and spectral analysis are $\sin^2 2\theta_{13} = 0.0896 \pm 0.0048(\rm stat) \pm 0.0047(\rm syst)$ and $|\Delta m_{ee}^2| = [2.68 \pm 0.12({\rm stat}) \pm 0.07({\rm syst})]\times 10^{-3}$~eV$^2$ with $\chi^2 /NDF = 47.4/66$, where $NDF$ is the number of degrees of freedom.
The statistical errors are reduced almost by a factor of two with respect to the previous measurement~\cite{RENO-spect1, RENO-spect2}.
The systematic error of $|\Delta m_{ee}^2|$ is significantly reduced by 45\% while that of $\sin^2 2\theta_{13}$ is reduced by 15\%.
The background uncertainty contributes $\pm$0.0021 to the systematic error of $\sin^2 2\theta_{13}$. The dominant contribution to the systematic error is due to the uncertainties of reactor flux ($\pm$0.0032) and detection efficiency ($\pm$0.0032). The systematic error of $|\Delta m_{ee}^2|$ comes mostly from the energy scale uncertainty.
The measured value of $|\Delta m_{ee}^2|$ corresponds to $|\Delta m_{32}^2| = (2.63 \pm 0.14)\times 10^{-3}$~eV$^2$ for the normal neutrino mass ordering and $(2.73 \pm 0.14)\times 10^{-3}$~eV$^2$ for the inverted neutrino mass ordering, using measured oscillation parameters of $\sin^2 \theta_{12} = 0.307 \pm 0.013$ and $\Delta m_{21}^2 = (7.53 \pm 0.18)\times 10^{-5}$~eV$^2$~\cite{PDG}.

\begin{figure}[hbt]
\begin{center}
\includegraphics[width=0.47\textwidth]{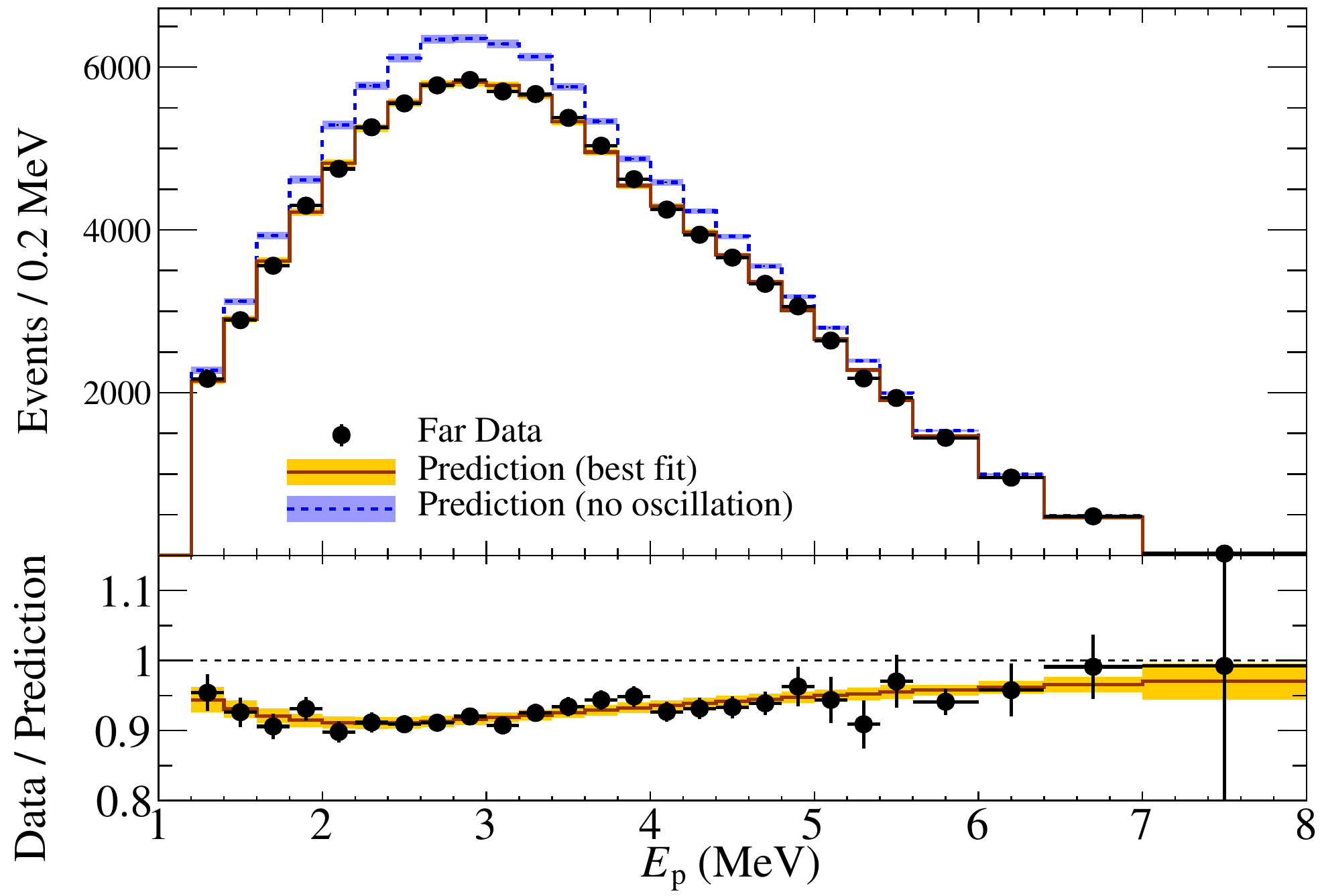}
\caption{Top: comparison of the observed IBD prompt spectrum in the far detector (dots) with the no-oscillation prediction (blue shaded histogram) obtained from the measurement in the near detector. The prediction from the best-fit oscilation parameters is also shown (yellow shaded histogram). Both blue and yellow bands represent uncertainties. Bottom: ratio of IBD events measured in the far detector to the no-oscillation prediction (dots) and the ratio from the MC simulation with best-fit results folded in (shaded band). Errors include the statistical and background subtraction uncertainties.}
\label{fig:spectra-oscillation-fit}
\end{center}
\end{figure}

\begin{figure}[hbt]
\begin{center}
\includegraphics[width=0.47\textwidth]{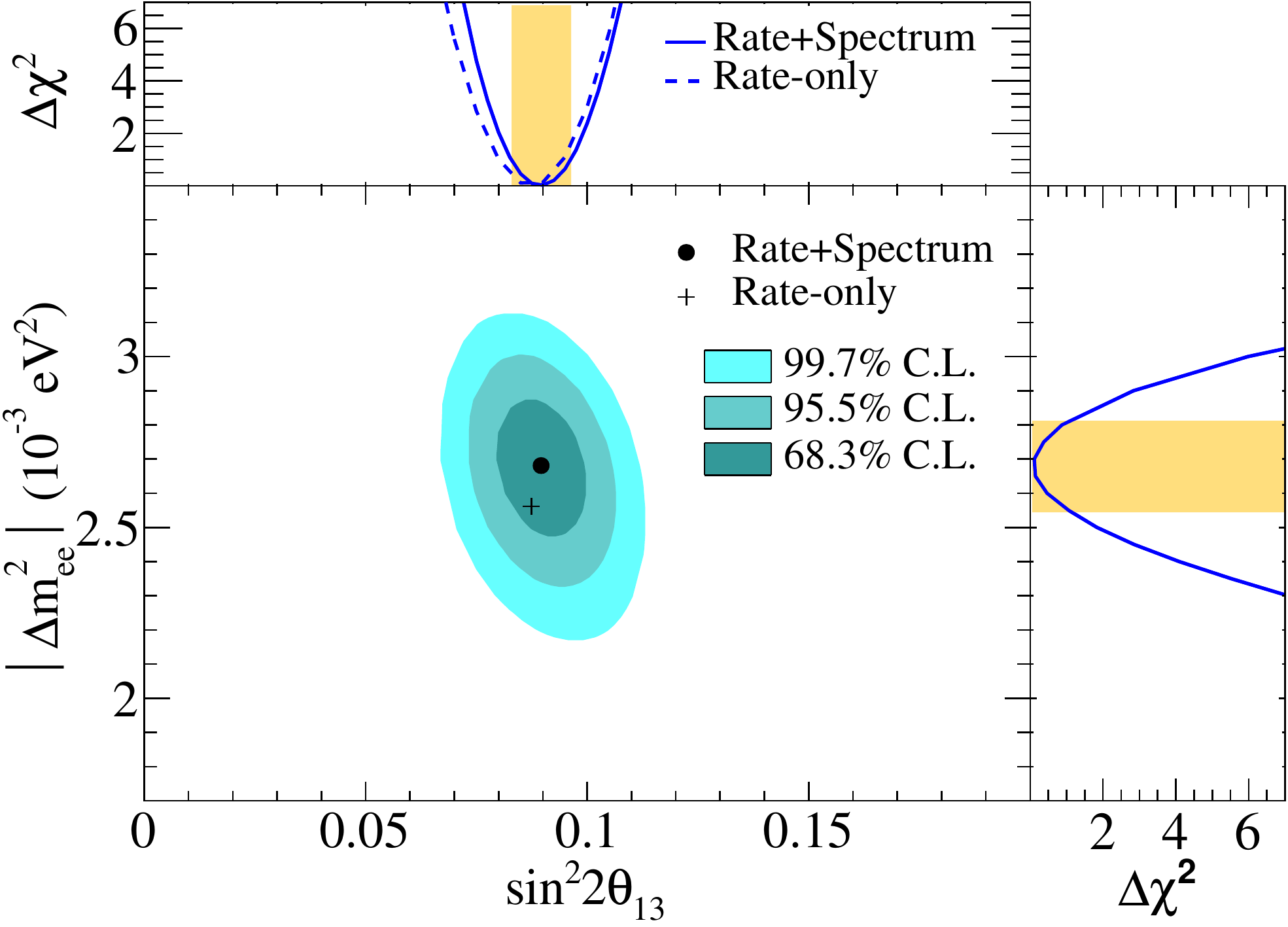}
\caption{Allowed regions of  68.3, 95.5, and 99.7\% C.L. in the $|\Delta m_{ee}^2|$ vs $\sin^2 2\theta_{13}$ plane. The best-fit values are given by the black dot. The $\Delta \chi^2$ distributions for $\sin^2 2\theta_{13}$ (top) and $|\Delta m_{ee}^2|$ (right) are also shown with an $1 \sigma$ band. The rate-only result for $\sin^2 2\theta_{13}$ is shown by the cross.} 
\label{fig:contour-allowed}
\end{center}
\end{figure}

Figure \ref{fig:spectra-oscillation-fit} shows the background-subtracted, IBD prompt energy spectrum at the far detector compared to the one expected with no oscillation and the one expected with the best-fit oscillation parameters at the far detector. The expected spectrum with no oscillation at the far detector is obtained by weighting the measured spectrum at the near detector with no-oscillation assumptions in order to include the 5-MeV excess.
The expected spectrum with the best-fit oscillation parameters is obtained by applying the measured values of $\sin^2 2\theta_{13}$ and $|\Delta m_{ee}^2|$ to the one expected with no oscillation at the far detector.
The observed spectrum at the far detector shows a clear energy dependent disappearance of reactor $\overline{\nu}_e$ consistent with neutrino oscillations. 
Figure \ref{fig:contour-allowed} shows 68.3, 95.5, and 99.7\% C.L. allowed regions for the neutrino oscillation parameters $|\Delta m_{ee}^2|$ and $\sin^2 2\theta_{13}$. 
     
\begin{figure}[hbt]
\begin{center}
\includegraphics[width=0.47\textwidth]{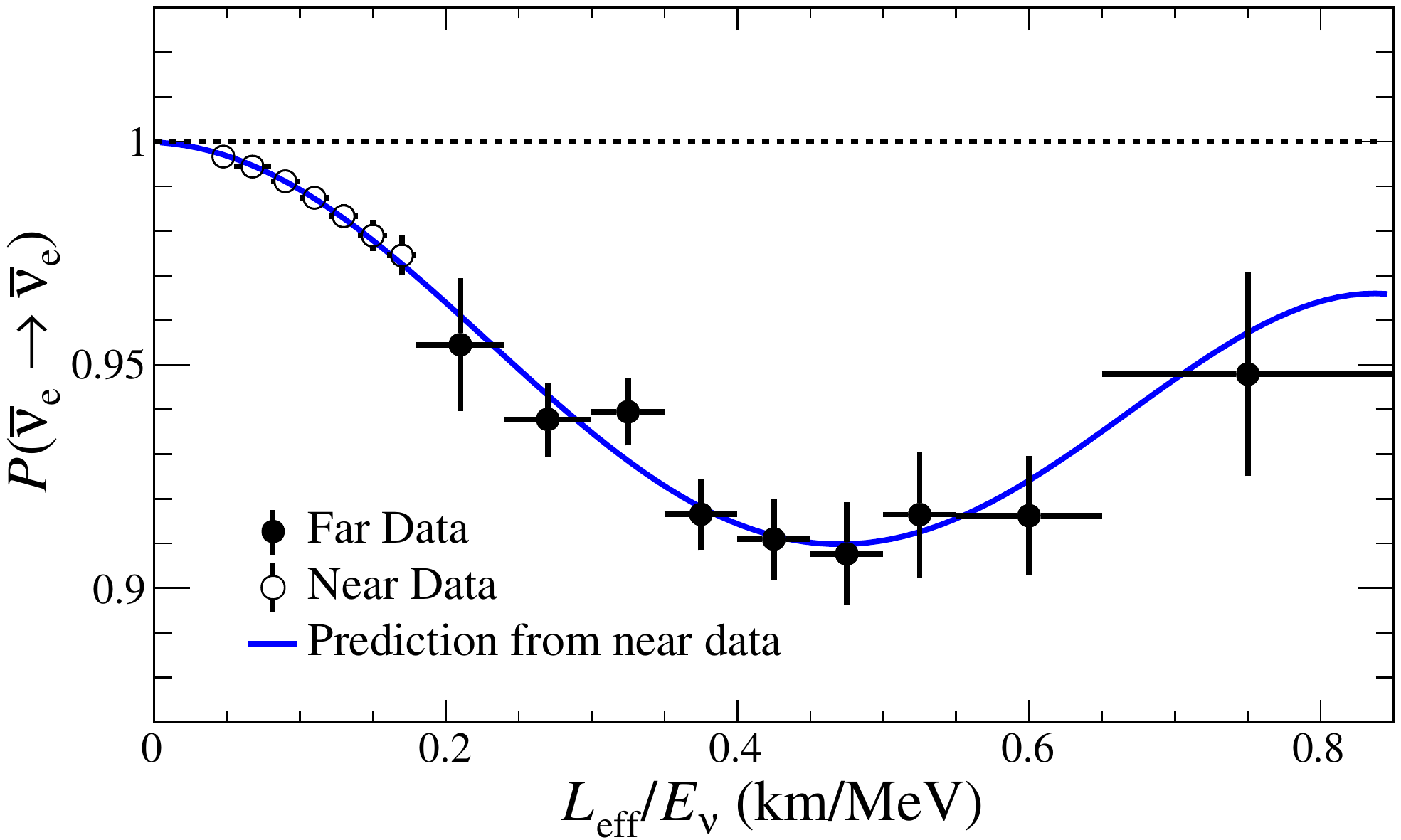}
\caption{Measured reactor $\overline{\nu}_e$ survival probability in the far detector as a function of $L_{\rm eff}/E_{\nu}$. The curve is a predicted survival probability, obtained from the observed probability in the near detector, for the best-fit values of $|\Delta m_{ee}^2|$ and $\sin^2 2\theta_{13}$. The $L_{\rm eff}/E_{\nu}$ value of each data point is given by the average of the counts in each bin.}
\label{fig:baseline-energy}
\end{center}
\end{figure}
The survival probability of reactor $\overline{\nu}_e$ is a function of a baseline over neutrino energy.
Because of having multiple reactors as neutrino sources, an effective baseline $L_{\rm eff}$ is defined by the reactor-detector distance weighted by the IBD event rate from each reactor.
Figure \ref{fig:baseline-energy} 
shows the measured survival probability of reactor $\overline{\nu}_e$ in the far detector as a function of an effective baseline $L_{\rm eff}$ over $\overline{\nu}_e$ energy $E_{\nu}$. The observed $L_{\rm eff}/E_{\nu}$ distribution is obtained by summing up the daily distributions weighted by a daily IBD rate.
The measured survival probability is obtained by the ratio of the observed IBD events to the expected ones with no oscillation in each bin of $L_{\rm eff}/E_{\nu}$.
A predicted survival probability is obtained from the observed probability distribution in the near detector and the best-fit oscillation values. 
A clear $L_{\rm eff}/E_{\nu}$-dependent disappearance of reactor $\overline{\nu}_e$ is observed and demonstrates the periodic feature of neutrino oscillation.

\begin{figure}[hbt]
\begin{center}
\includegraphics[width=0.47\textwidth]{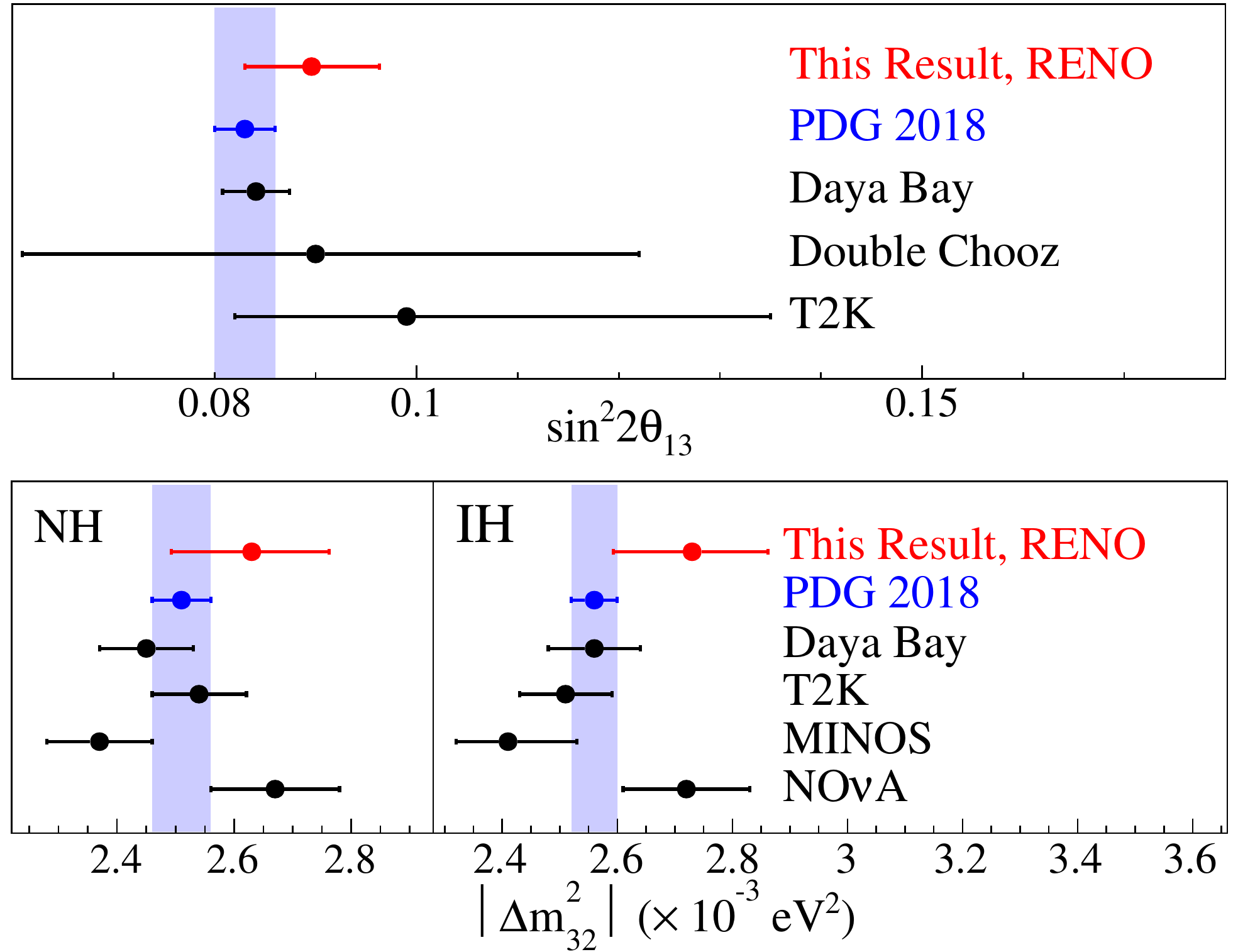}
\caption{Comparison of experimental results on $\sin^2 2\theta_{13}$ and $|\Delta m_{32}^2|$. The world average values and the experimental results of Daya Bay~\cite{DB-results}, Double Chooz~\cite{DC-results}, T2K~\cite{T2K-results}, MINOS~\cite{MINOS-results}, and NO$\nu$A~\cite{NOvA-results} are taken from the Ref.~\cite{PDG}.}
\label{fig:comparison}
\end{center}
\end{figure}

In summary, RENO has observed clear energy dependent disappearance of reactor $\overline{\nu}_e$ using two identical detectors, and obtains $\sin^2 2\theta_{13} = 0.0896 \pm 0.0068 $ and $|\Delta m_{ee}^2| = (2.68 \pm 0.14 )\times 10^{-3}$~eV$^2$ based on the measured periodic disappearance expected from neutrino oscillations. 
With the increased statistics of the 2\,200 day data sample and the reduced background rates, RENO has produced a precise measurement of the reactor $\overline{\nu}_e$ oscillation amplitude and frequency. The measured uncertainty is reduced from 0.0100 to 0.0068 for $\sin^2 2\theta_{13}$ and from $0.25\times 10^{-3}$~eV$^2$ to $0.14\times 10^{-3}$~eV$^2$ for $|\Delta m_{ee}^2|$, relative to the previous measurement~\cite{RENO-spect1, RENO-spect2}.
 The RENO's measured values of $\sin^2 2\theta_{13}$ and $|\Delta m_{32}^2|$ are compared with other experimental results in Fig. \ref{fig:comparison}.
It would provide an important information on the determination of the leptonic CP phase if combined with a result of an accelerator neutrino beam experiment. 

The RENO experiment is supported by the National Research Foundation of Korea (NRF) Grant No. 2009-0083526 funded by the Korea Ministry of Science and ICT. Some of us have been supported by a fund from the BK21 of the NRF and Institute for Basic Sicence grant No. IBS-R017-G1-2018-a00.
We gratefully acknowledge the cooperation of the Hanbit Nuclear Power Site and the Korea Hydro and Nuclear Power Co., Ltd. (KHNP). 
We thank KISTI for providing computing and network resources through GSDC, and all the technical and administrative people who greatly helped in making this experiment possible.

\end{document}